\documentclass[aps,prb,preprint,groupedaddress]{revtex4-1}
\usepackage{graphicx,amsmath}
\usepackage{epstopdf}

\begin{document}


\title{Origin of Multiferroicity in MnWO$_4$}


\author{I. V. Solovyev}
\email{SOLOVYEV.Igor@nims.go.jp}
\affiliation{Computational Materials Science Unit,
National Institute for Materials Science, 1-2-1 Sengen, Tsukuba,
Ibaraki 305-0047, Japan}


\date{\today}

\begin{abstract}
MnWO$_4$ is regarded as a canonical example of multiferroic materials, where
the multiferroic activity is caused by a spin-spiral alignment.
We argue that, in reality, MnWO$_4$ has two sources of the spin-spirality,
which conflict with each other.
One is the Dzyaloshinskii-Moriya (DM) interactions, reflecting the
$P2/c$ symmetry of the lattice.
The $P2/c$ structure of MnWO$_4$ has an inversion center, that connects two Mn sublattices.
Therefore, from the viewpoint of DM interactions, different Mn sublattices are expected
to have \textit{opposite} spin chirality.
Another source is competing isotropic exchange interactions, which tend to form a
spin-spiral texture with
\textit{the same} chirality in both magnetic sublattices.
Thus, there is a conflict between DM and isotropic exchange
interactions, which makes these two sublattices inequivalent and, therefore,
breaks the inversion symmetry.
Our theoretical
analysis is based on the low-energy model, derived from the first-principles electronic
structure calculations.
\end{abstract}


\maketitle

\section{\label{sec:intro} Introduction}

  The idea of breaking the inversion symmetry by some complex magnetic
order has attracted a great deal of attention. It gives rise to the phenomenon of
improper multiferroicity, where the ferroelectric (FE) polarization is
induced by the magnetic order and, therefore, can be controlled by the magnetic field.
Alternatively, one can change the magnetic structure by applying the electric field.
This makes multiferroic materials promising for the creation of the new generation of
electronic devices.\cite{MF_review,spiral_review}

  One of magnetic textures, that breaks the inversion symmetry, is the spin spiral.
Moreover, this property is universal, in a sense that it can take place in all types
of magnetic compounds, irrespectively of their symmetry, as long as the
the spin-spiral order is established.\cite{Sandratskii} Probably, this is the reason why
the idea of the spin-spiral alignment became so popular in the field of multiferroics:
many properties of such compounds are interpreted from the viewpoint of
spin-spiral order,\cite{KNB,Mostovoy,Sergienko} and the search for new multiferroic
materials is frequently conducted around those with the spin-spiral texture.\cite{spiral_review}

  Nevertheless, the spin-spiral alignment alone is insufficient for having a FE activity:
although the spin-spiral order breaks the inversion symmetry, one can always find
some appropriate uniform rotation of spins, which, after combining with the spatial inversion,
will transform the spin-spiral texture to itself (see Appendix~\ref{sec:appendix}). The
FE polarization in such a situation will be equal to zero. In order to make it finite
(and in order to fully break the inversion symmetry), one should ban the possibility of uniform rotations
in the system of spins.
This can be done by the relativistic spin-orbit (SO) interaction, and today it is commonly
accepted in the theory of multiferroic materials that the spin-spiral alignment
should be always supplemented with the SO coupling.\cite{KNB,Sergienko}
However, as long as the SO coupling is involved, the spin-spiral texture becomes deformed.
Then, we face the question whether the FE activity is caused by the spin-spiral order itself
or by the deviation from it.\cite{antispiral}
The answer to this question is very important for understanding the origin of multiferroicity.

  MnWO$_4$ is an important material in the field of multiferroics.
Although the transition temperature
to the multiferroic phase is low and
the FE polarization is weak, the material is fundamentally important, because: (i) its multiferroic
phase has a spin-spiral texture (or is believed to be the spin spiral);
(ii) the relative directions of the spin-spiral propagation vector,
the spin rotation axis, and the FE polarization seem to be consistent with predictions based
on the spin-spiral theory.\cite{Taniguchi,Arkenbout,Heyer} Thus, MnWO$_4$ is regarded as one of
successful manifestations of the spin-spiral theory and, actually, the discovery of the FE activity in MnWO$_4$
was guided by this theory.

  More specifically, MnWO$_4$ crystallizes in the monoclinic $P2/c$ structure.
It exhibits three successive antiferromagnetic (AFM)
transitions: at $T_{\rm N3}$$\sim 13.5$ K, $T_{\rm N2}$$\sim 12.5$ K, and $T_{\rm N1}$$\sim 7.6$ K.
The AF1 phase, realized below $T_{\rm N1}$, is nearly collinear and the directions of spins
alternate along the monoclinic ${\bf b}$ axis as $\uparrow \uparrow \downarrow \downarrow$
(the reason why the AF1 phase is also called as the `$\uparrow \uparrow \downarrow \downarrow$ phase').
The AF2 phase ($T_{\rm N1}$$<T$$<$$T_{\rm N2}$) is incommensurate
and is typically ascribed to the spin spiral with the
propagation vector
${\bf q}_{\rm AF2} = (-$$0.214,1/2,0.457)$ (measured in units of reciprocal lattice translations).
The spin moments lie in the plane formed by the easy magnetization direction in the plane
$\boldsymbol{ac}$ and the $\boldsymbol{b}$ axis.
The AF3 phase ($T_{\rm N2}$$<$$T$$<$$T_{\rm N3}$) is characterized by a collinear sinusoidally
modulated AFM order with the same ${\bf q}$.
The AF2 phase is ferroelectric. The polarization vector is
parallel to the monoclinic $\boldsymbol{b}$ axis: ${\bf P} = (0,P_b,0)$.
$P_b$ is relatively small: it takes a maximal value ($\sim$ 50 $\mu$C/m$^2$) at around $T_{\rm N1}$, and
then monotonously decreases and vanishes at around $T_{\rm N2}$.\cite{Taniguchi}
Other phases display no sign of FE activity.

  In this work, we will argue that the multiferroicity in MnWO$_4$ has a more complex origin. Namely, we will show that:
(i) It is not sufficient to have a simple spin-spiral texture; (ii) The spin spiral should be deformed by some
conflicting interactions, existing in the system. In MnWO$_4$, these are Dzyaloshinskii-Moriya (DM) interactions
and isotropic exchange interactions. The conflict of these two interactions breaks the inversion
symmetry and gives rise to the FE activity in the AF2 phase.

\section{\label{sec:method} Method}

  We follow the same strategy as in our previous publications,
devoted to multiferroic manganites.\cite{antispiral} Our basic idea is to construct a realistic
low-energy (Hubbard-type) model for the Mn $3d$ bands, which would be sufficient for describing basic
magnetic properties of MnWO$_4$ at a semi-quantitative level. The model is constructed in the basis of
Wannier orbitals, starting from the electronic structure calculations
in the local density approximation (LDA) with and without the relativistic SO interaction. The details
can be found in the review article (Ref.~\onlinecite{review2008}).
All calculations have been performed using experimental parameters of the
crystal structure.\cite{Lautenschlaeger}
Since heavy W atoms is an important source of relativistic effects,
the SO coupling was included already at the level of ordinary LDA calculations, and thus
obtained electronic structure was used as the starting point for the
construction of the low-energy model.
Thus, although the W sites were not explicitly included to the model, the relativistic effects,
associated with these sites, were allowed to contribute to the dispersion
of the Mn $3d$ bands and, in this way, were taken into consideration during the analysis of
the low-energy model. All obtained parameters of the model Hamiltonian
are collected in Ref.~\onlinecite{SM}. After the construction, the model was solved in the
mean-field Hartee-Fock (HF) approximation.

  We would also like to briefly comment on merits and demerits of our technique in comparison with
the first-principles approach:

  (i) It is certainly true that the construction of the model Hamiltonian is based on some
additional approximations,\cite{review2008} and is sometimes regarded as a step back
in comparison with first-principles calculations. On the other hand, it would not be right to think
that the first-principles calculations for the transition-metal oxides are free of any approximations.
The necessity to treat the problem of on-site Coulomb correlations, which is frequently formulated
in terms of the LDA$+$$U$ approach (with some phenomenological correction ``$+$$U$'', borrowed from the
Hubbard model) make it similar to the model approach: it also relies on additional approximations,
although of a different type. The typical approximations are the
choice of the parameter $U$ and the form of the double-counting term, which are both
ill-defined in LDA$+$$U$.\cite{PRB98} Thus, at the present stage, it would not be right
to say that one techniques is more superior than other: rather, they provide a complementary information
for the analysis of material properties of the transition-metal oxides.
As we will see below, results of our model analysis are well consistent with available
first-principles calculations for MnWO$_4$.\cite{Tian,Shanavas}

  (ii) The search of the ground state for noncollinear magnetic textures can be very
time consuming, even at the level of mean-field HF approximation: the rotations of magnetic moments
towards new equilibrium positions after including the SO interactions can be very slow and require tens of thousands of iterations,
as in the case of MnWO$_4$. In such a situation, the model Hamiltonian approach is very useful,
because it allows us to find a fully self-consistent
magnetic structure, which is not always accessible for
the ordinary first-principles calculations.\cite{antispiral}

 (iii) The model Hamiltonian approach can be very useful for the analysis and interpretation of results
of complex electronic structure calculations and the experimental data. Particularly, in this work it will
help us to elucidate the microscopic origin of the magnetic inversion symmetry breaking in MnWO$_4$.
We will also show that the experimental behavior of long-range magnetic interactions
in MnWO$_4$ (Ref.~\onlinecite{Ye}) can be naturally understood in the framework of the superexchange (SE)
theory and reflects similar behavior of the transfer integrals, derived for the low-energy model.

\section{\label{sec:results} Results and discussions}

\begin{figure}[h!]
\begin{center}
\includegraphics[width=7.1cm]{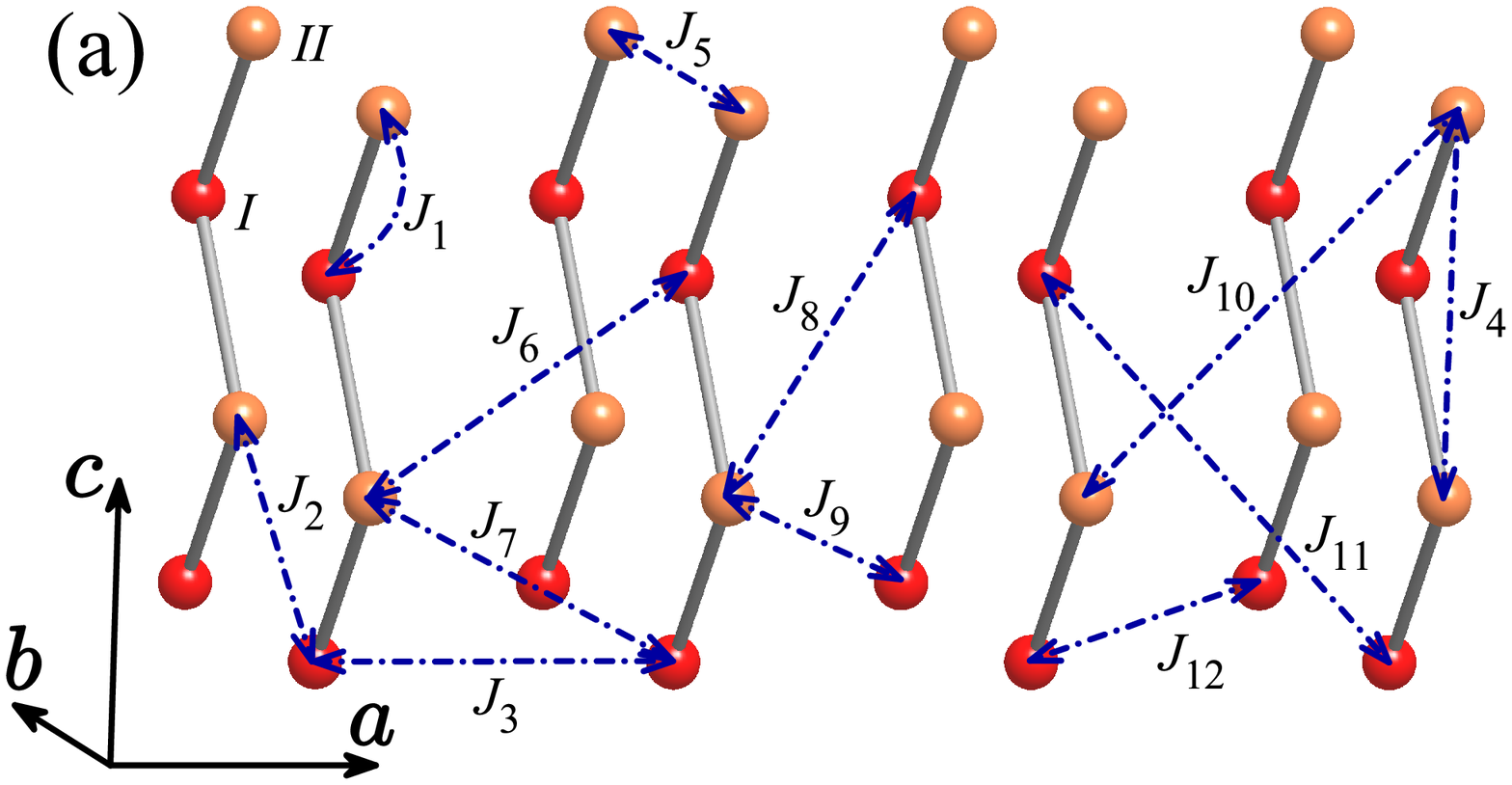}
\includegraphics[width=5.8cm]{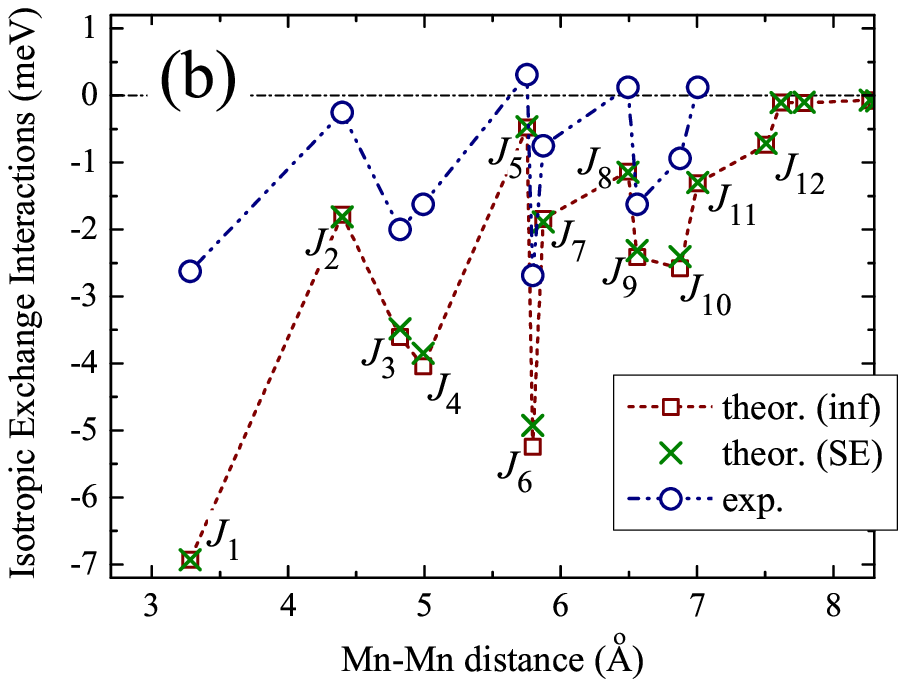}
\end{center}
\caption{\label{fig.J}
(Color online) (a) Lattice of Mn sites with the notations of
isotropic exchange interactions
(numbered in the increasing order of interatomic distances).
Note that the $P2/c$ structure of
MnWO$_4$ has two Mn sublattices, which are shown by different colors
and denoted as `$I$' and `$II$', respectively. These sublattices are
transformed to each other by the inversion operation.
(b) Distance-dependence of isotropic exchange interactions:
results of calculations, using the theory of infinitesimal
spin rotations near the ferromagnetic state (denoted as `inf') and the
theory of superexchange interactions with $\Delta_{\rm ex} = 5$ eV (denoted as `SE'), in comparison with
the experimental data from Ref.~\protect\onlinecite{Ye}.
}
\end{figure}

  First, we solve the low-energy model in the HF approximation,
by assuming the collinear ferromagnetic (FM) alignment of spins, and drive parameters of
isotropic exchange interactions. For these purposes we employ the theory of
infinitesimal spin rotations.\cite{review2008,Liechtenstein}
The procedure corresponds to the local mapping of the change of the one-electron energy onto the spin Hamiltonian
of the form
${\cal H}_S = -\sum_{\langle ij \rangle} J_{ij} {\bf e}_i \cdot {\bf e}_j$, where
${\bf e}_i$ is the \textit{direction} of spin at the site $i$ and the
summation run over \textit{inequivalent pairs} of sites.
The results of these calculations are explained in Fig.~\ref{fig.J}.
Alternatively, one can apply the theory of superexchange (SE) interactions,
by considering the energy gain caused by virtual hoppings ($t_{ij}^{mm'}$) in
the second order of perturbation theory.\cite{PWA}
For the $d^5$ configuration of Mn$^{2+}$,
this expression is extremely simple:
$J_{ij} = -$$\sum_{mm'} t_{ij}^{mm'} t_{ji}^{m'm}/\Delta_{\rm ex}$, where $\Delta_{\rm ex}$
is the intraatomic splitting between the majority- and minority-spin states and the summation runs
over all $3d$ orbitals: $m$ $(m') =$ $xy$, $yz$, $3z^2$$-$$r^2$, $zx$, and $x^2$$-$$y^2$.
Thus, all SE interactions are expected to be antiferromagnetic.
Then, by taking $\Delta_{\rm ex} = 5$ eV and using the values of
$t_{ij}^{mm'}$ obtained for the low-energy model (Ref.~\onlinecite{SM}), one can find that the SE theory
excellently reproduces the
results of the more general theory of infinitesimal spin rotations.
Moreover, the value of $\Delta_{\rm ex} = 5$ eV
is well consistent with typical estimates in the framework of the Hubbard model, where
$\Delta_{\rm ex} \approx U$$+$$4J_{\rm H}$ and the parameters of averaged on-site Coulomb repulsion
$U$ and the exchange interaction $J_{\rm H}$,
derived for the low-energy model, are $1.8$ and $0.8$ eV, respectively.\cite{SM}
Thus, all estimates are consistent with each other. Therefore, we may conclude that:
(i) the physically relevant mechanism, responsible for isotropic exchange interactions
in MnWO$_4$, is the superexchange; and (ii) the distance-dependence of
$J_{ij}$ reflects the behavior of the transfer integrals $t_{ij}^{mm'}$.

  Our calculations capture main details of the experimental
distance-dependence of $J_{ij}$ (Fig.~\ref{fig.J}).\cite{Ye} This is a very important finding, which means that
in our model we should be able to reproduce the correct
magnetic structures of MnWO$_4$.
The negative aspect is that our parameters $J_{ij}$ are `too antiferromagnetic' (AFM).
For example, using the values of $J_{ij}$ reported in Fig.~\ref{fig.J},
the Curie-Weiss temperature can be estimated as
$\theta_{\rm CW} \approx \sum_j J_{ij}/3 k_{\rm B}$.
It yields $\theta_{\rm CW}$$= -$$265$ K, which exceeds the experimental value
$-$$78$ K (Ref.~\onlinecite{Taniguchi}) by factor three.
The discrepancy cannot be resolved simply by changing the value of the on-site Coulomb repulsion $U$,
which is frequently regarded as an ill-defined parameter. Particularly, in order to explain the existence of
FM interactions, which were observed experimentally in some of the bonds,\cite{Ye} we need an additional
mechanism on the top of our model. Such a mechanism can be related to the
magnetic polarization of the oxygen band, similar to orthorhombic manganites.\cite{JPSJ}
The first-principles GGA$+$$U$ calculations
(where GGA stands for the generalized gradient approximation -- an extension of LDA)
also substantially overestimate $| J_{ij} |$ and $| \theta_{\rm CW} |$.\cite{Tian}
Although these calculations take into account the effect of the oxygen band,
the disagreement is probably caused by additional approximations for the double-counting energy,
which is ill-defined in GGA$+$$U$.\cite{PRB98} Finally, the experimental parameters $J_{ij}$
themselves may be sensitive to the details of fitting of the spin-wave dispersion. In this respect,
we would like to mention that earlier experimental data were interpreted
in terms of rather different set of parameters $J_{ij}$.\cite{Ehrenberg}
Thus, at the present stage, taking into account the complexity of
the problem of interatomic magnetic interactions in MnWO$_4$,
the agreement with the experimental data in Fig.~\ref{fig.J} can be regarded as satisfactory.

\begin{figure}[h!]
\begin{center}
\includegraphics[width=10cm]{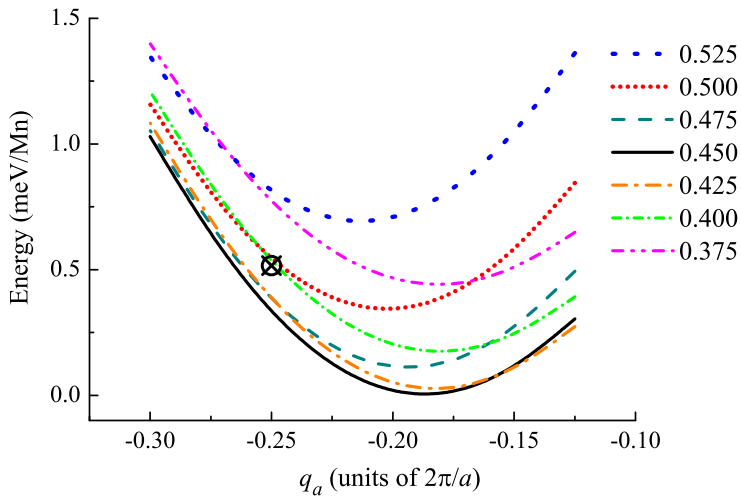}
\end{center}
\caption{\label{fig.Eq}
(Color online) Results of Hartree-Fock calculations
for the spin-spiral states with ${\bf q} = (q_a,1/2,q_c)$: total energies versus $q_a$ for different values of $q_c$
(in units of $2\pi/c$).
The energy of the collinear $\uparrow \uparrow \downarrow \downarrow$ state is marked by the symbol $\otimes$.
}
\end{figure}

  Then, we search for the magnetic ground state of MnWO$_4$ without relativistic SO coupling.
For these purposes we employ the spin-spiral formalism, which is based on the generalized Bloch
theorem.\cite{Sandratskii} The idea is to combine the lattice translations with
rotations in the spin subspace. It allows us to treat the incommensurate magnetic textures
with an arbitrary vector ${\bf q}=(q_a,1/2,q_c)$ in the reciprocal space
such as if the magnetic primitive cell was of the same size as the
crystallographic one.\cite{Sandratskii} The results of these
calculations, which were also performed in the HF approximation
for the low-energy model, are summarized in Fig.~\ref{fig.Eq}.
For ${\bf q} = (-$$1/4,1/2,1/4)$, the
collinear $\uparrow \uparrow \downarrow \downarrow$ and noncollinear spin-spiral states
are nearly degenerate. Nevertheless, the total energy of the spin spiral continues to decrease
with the increase of $q_a$. Therefore, the `nonrelativistic' ground state of MnWO$_4$ is expected to be
the spin spiral with ${\bf q}_{\rm NR} \approx (-$$0.175,1/2,0.45)$, which is lower in energy than the
$\uparrow \uparrow \downarrow \downarrow$ state
by about $0.5$ meV/Mn.
Moreover, none of these states is ferroelectric: although the spin spiral breaks
the inversion operation, the latter can be always combined with an appropriate spin rotation, which
will transform
the inverted spin spiral to the original one. Thus,
the FE polarization will vanish (see Appendix~\ref{sec:appendix}).

  Therefore, we should find the answer to the following questions: (i) Which interaction stabilizes
the AF1 structure and make it the ground state of MnWO$_4$
at low $T$? (ii) What is the mechanism of the inversion-symmetry breaking in
the AF2 state, which yields finite $P_b$?
For these purposes we consider the relativistic SO interaction.
It gives rise to such important ingredients as the single-ion anisotropy and DM interactions.

  The effect of single-ion anisotropy on the magnetic texture of MnWO$_4$ can be studied by
enforcing the atomic limit and setting all transfer integrals equal to zero.\cite{PRB12}
Generally, the single-ion anisotropy is expected to be small for the nearly spherical
$d^5$ configuration of Mn$^{2+}$.
Since the symmetry operation $\{ C_b^2|\boldsymbol{c}/2 \}$ of the $P2/c$ space group transforms each Mn site to itself
(here, $C_b^2$ stands for the $180^\circ$ rotation around the $\boldsymbol{b}$ axis and $\boldsymbol{c}/2$
is the additional translation),
the single-ion anisotropy will tend to align the spins either parallel
to the $\boldsymbol{b}$ axis ($|| \boldsymbol{b}$)
or in the $\boldsymbol{ac}$ plane ($\in \boldsymbol{ac}$). In the latter case, the spins are canted
off the $\boldsymbol{a}$ axis by about $41^\circ$,
which is close to the experimental value of $35^\circ$.\cite{Taniguchi}
The easy magnetization direction corresponds to the in-plane configuration. However, the energy difference between
hard ($|| \boldsymbol{b}$) and easy ($\in \boldsymbol{ac}$) magnetization
directions is small (less that $0.05$ meV/Mn), so that in the spin-spiral texture it
can be overcome by the energy gain caused by the isotropic exchange interactions.

  The $P2/c$ structure of MnWO$_4$ has two Mn-sublattices (denoted as `$I$' and `$II$' in Fig.~\ref{fig.J}),
which are transformed to each other by the inversion operation.
Therefore, the DM interactions will obey the following symmetry rules:\cite{DM}
\begin{itemize}
\item[(i)]
all interactions between different sublattices are equal to zero;
\item[(ii)]
the DM interactions in the sublattices `$I$' and `$II$' are related by the identity
${\bf d}^{I}_{\boldsymbol{R}} = {\bf d}^{II}_{-\boldsymbol{R}} = -$${\bf d}^{II}_{\boldsymbol{R}}$,
for any translation $\boldsymbol{R}$, connecting the sites within one Mn sublattice
(note that ${\bf d}$ is the axial vector).
\end{itemize}
From now on, it is more convenient to use the
extended notations and recall that each site of the lattice $i$ is specified by its
position $\boldsymbol{\tau}$ in the primitive cell and the translation $\boldsymbol{R}$.

  Moreover, due to the symmetry operation $\{ C_b^2|\boldsymbol{c}/2 \}$,
all DM vectors lie in the $\boldsymbol{ac}$ plane, i.e., similar to the easy magnetization
direction, obtained from the analysis of
the single-ion anisotropy.
The parameters of DM interactions can be evaluated by considering the mixed type of the perturbation theory
expansion with respect to the SO coupling and infinitesimal rotations of
spins near the FM state with subsequent mapping of obtained results for the change of the
one-electron energy onto the spin Hamiltonian of the form
$\sum_{\langle \boldsymbol{R} \boldsymbol{R}' \rangle } {\bf d}_{\boldsymbol{R} - \boldsymbol{R}'}
[{\bf e}_{\boldsymbol{R}} \times {\bf e}_{\boldsymbol{R}'} ]$, constructed
for each magnetic sublattice.\cite{PRL96} It yields the following parameters of DM interactions (in meV):
${\bf d}^{I}_{\boldsymbol{a}} = (-$$0.01,0,0.01)$, ${\bf d}^{I}_{\boldsymbol{c}} = (-$$0.02,0,0.02)$,
and ${\bf d}^{I}_{\boldsymbol{a}+\boldsymbol{c}} = (0.01,0,-$$0.01)$. Interactions in other bonds are
considerably smaller.

\begin{figure}[h!]
\begin{center}
\includegraphics[width=10cm]{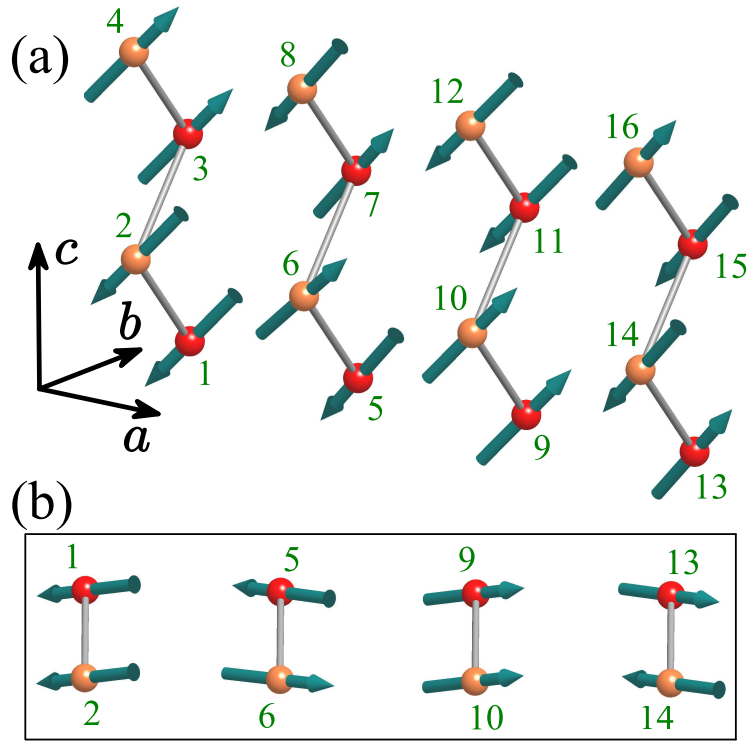}
\end{center}
\caption{\label{fig.uudd}
(Color online) (a) The
magnetic structure AF1, as obtained in the mean-field HF calculations for the
low-energy model with relativistic SO interaction, and (b) its projection onto the $\boldsymbol{ab}$ plane.
}
\end{figure}

  The equilibrium magnetic structure, obtained
after switching on the relativistic
SO interaction
in the collinear
AF1 state, is explained in Fig.~\ref{fig.uudd}.
Due to the single-ion anisotropy,
all spins are
confined mainly in the $\boldsymbol{ac}$ plane and
canted off the $\boldsymbol{a}$ axis by about $46^\circ$
(which is close to $41^\circ$, obtained in the atomic limit).
The DM interactions yield an additional canting
out of the $\boldsymbol{ac}$ plane. The corresponding rotational force ${\bf f}$, acting at
some Mn spin from its neighboring sites in the same
sublattice, is given by
${\bf f} = \sum_{\boldsymbol{R}} [ {\bf d}_{\boldsymbol{R}} \times {\bf e}_{\boldsymbol{R}} ]$.
Since ${\bf d}_{-\boldsymbol{R}} = -$${\bf d}_{\boldsymbol{R}}$,
in order to contribute to ${\bf f}$, the directions of neighboring spins
should also satisfy the condition ${\bf e}_{-\boldsymbol{R}} = -$${\bf e}_{\boldsymbol{R}}$.
In the $\uparrow \uparrow \downarrow \downarrow$ texture, such a situation takes place for
$\boldsymbol{R} =$ $\boldsymbol{a}$ and
$\boldsymbol{a}$$+$$\boldsymbol{c}$, but not for
$\boldsymbol{R} =$ $\boldsymbol{c}$.
Moreover, since both ${\bf d}_{\boldsymbol{R}}$ and
${\bf e}_{\boldsymbol{R}}$ lie in the $\boldsymbol{ac}$ plane, the force
${\bf f}$ is parallel to the $\boldsymbol{b}$ axis.
Finally, due to the alternation of ${\bf e}_{\boldsymbol{R}}$
in the $\uparrow \uparrow \downarrow \downarrow$ texture,
the directions of such forces will also alternate. All in all, this explains fine details of the
magnetic structure in Fig.~\ref{fig.uudd}b.
Very importantly, the isotropic exchange interactions, single-ion anisotropy, and DM interactions
in the AF1 phase
do not conflict with each other in a sense that all of them,
even jointly, continue to respect the crystallographic $P2/c$ symmetry.
Therefore, the inversion symmetry is preserved and the polarization is equal to zero, even
after including the SO interaction.

\begin{figure}[h!]
\begin{center}
\includegraphics[width=15cm]{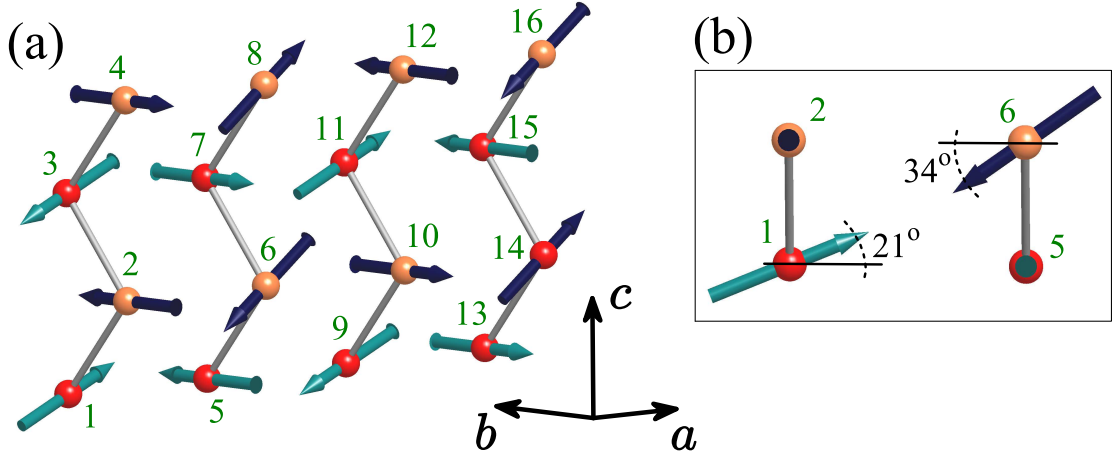}
\end{center}
\caption{\label{fig.spiral4}
(Color online) (a) The deformed spin-spiral
texture, as obtained in the mean-field HF calculations for the
low-energy model with relativistic SO interaction for
${\bf q} = (-$$1/4,1/2,1/2)$, and (b) its projection onto the $\boldsymbol{ac}$ plane.
}
\end{figure}

  The effect of SO interaction on the spins-spiral alignment can be best understood for
${\bf q} = (-$$1/4,1/2,1/2)$ (see Fig.~\ref{fig.spiral4}).
In this structure,
the easy magnetization direction (${\bf e} \in \boldsymbol{ac}$) at certain
Mn-site alternates with the hard magnetization direction (${\bf e} || \boldsymbol{b}$)
at its neighboring sites. Thus, there is a
conflict between isotropic exchange interactions and the single-ion anisotropy. However, since the latter is small,
this magnetic structure can be realized at a low energy cost.
Then, due to the DM interactions,
each site with ${\bf e} || \boldsymbol{b}$ will experience a force from its
neighboring sites with ${\bf e} \in \boldsymbol{ac}$. Since ${\bf d} \in \boldsymbol{ac}$,
this force will be parallel to the
$\boldsymbol{b}$ axis. Alternatively, each site with ${\bf e} \in \boldsymbol{ac}$
will experience the rotational force from its neighboring sites with ${\bf e} || \boldsymbol{b}$.
Since ${\bf d} \in \boldsymbol{ac}$,
this force will also lie in the $\boldsymbol{ac}$ plane.

  Furthermore, since
${\bf d}^{I}_{\boldsymbol{R}} = -$${\bf d}^{II}_{\boldsymbol{R}}$,
the forces in different sublattices will act
\textit{in the opposite directions}.
This means that, if some DM vector ${\bf d}^{I}_{\boldsymbol{R}}$ in the first sublattice
will tend to support the right-handed spiral, its counterpart in the second sublattice
(${\bf d}^{II}_{\boldsymbol{R}}$) will support the left-handed spiral.
However, the spin-spiral texture, that is formed by the isotropic exchange interactions
$J_{ij}$, is always either right- or left-handed \textit{in both magnetic sublattices}.
Therefore, we have a conflict between isotropic and DM interactions: if in one magnetic
sublattice, the effect of these two interactions is added, in other magnetic
sublattice, it will be subtracted. All these tendencies are clearly seen
in fine details of the magnetic structure, obtained in the HF calculations with SO coupling (Fig.~\ref{fig.spiral4}).
First, the canting of spins, lying in the $\boldsymbol{ac}$ plane, appears to
be different for
the sublattices $I$ and $II$:
$21^\circ$ and $34^\circ$, respectively, relative to the $\boldsymbol{a}$ axis.
Second, as discussed above, the spins with ${\bf e} || \boldsymbol{b}$ will experience
the force ${\bf f}$, which is also parallel to $\boldsymbol{b}$. However,
since ${\bf d}^{I}_{\boldsymbol{R}} = -$${\bf d}^{II}_{\boldsymbol{R}}$,
the vectors ${\bf e}$ and ${\bf f}$ at the same site will be either parallel or antiparallel,
depending on the magnetic sublattice. Thus, the magnitude
of spin
magnetic moment with ${\bf e} || \boldsymbol{b}$
will also depend on the magnetic sublattice. Of course, this change of the
magnetic moments is small (about $0.01$ \%, according to our HF calculations).
Nevertheless, it does take place and is just another manifestation
of the conflict between isotropic and DM
interactions.

  Thus, the conflict between isotropic and DM interactions in the AF2 phase
makes the sublattices $I$ and $II$ inequivalent. Since in the
$P2/c$ structure, these two sublattices were connected
by the inversion operation,
their inequivalency means that the inversion symmetry is broken and the actual symmetry of the
AF2 phase is lower than $P2/c$.
Furthermore, the homogeneous spin-spiral alignment itself is
deformed by the DM interactions and,
strictly speaking, the AF2 phase is \textit{no longer the spin spiral}. The magnitude of this
deformation can be seen in the inset of Fig.~\ref{fig.spiral4}.

  Then, we discuss relative stability of different magnetic phases
with the SO interaction. In this case, the generalized Bloch
theorem is not applicable and there only possibility is to work with the
supercell geometry.\cite{Sandratskii} Therefore, we were able to consider
only solutions with ${\bf q} = (q_a,1/2,1/2)$ and the rational $q_a$$=$
$-$$1/3$, $-$$1/4$, and $-$$1/5$.
The results can be summarized as follows. Amongst AF2 states, the one
with $q_a$$=$ $-$$1/5$ has the lowest energy.
The state with $q_a$$=$ $-$$1/4$
is higher in energy by about 0.2 meV/Mn.
This behavior is similar to calculations without SO coupling (see Fig.~\ref{fig.Eq}).
Then, the AF2 state with $q_a$$=$ $-$$1/5$ appears to be nearly degenerate
with the AF1 state
(the energy difference is about $0.07$ meV/Mn, but the AF2 state is still low in energy).
Although these numerical values are probably on the verge of accuracy of our model analysis,
this tendency clearly shows that the SO interaction additionally stabilizes
AF1 state relative to the AF2 one. This seems to be reasonable: the conflict of
DM and isotropic exchange interactions, acting in the
opposite directions in one of the Mn sublattices, will
penalize the energy of the AF2 state. On the other hand, in the AF1 state, there is no such conflict
and the DM interactions will additionally minimize the energy of the $\uparrow \uparrow \downarrow \downarrow$ structure,
obtained without SO interaction. Apparently, the main reason why the energy of the AF2 phase is still
slightly lower than that of the AF1 phase is related to the fact that the isotropic exchange interactions
are overestimated in our approach (see Fig.~\ref{fig.J}) and, therefore, the effect of the SO coupling and
DM interactions on the magnetic structure is underestimated.

  Finally, we evaluate the value of the FE polarization in the AF2 phase, using the
Berry phase formalism,\cite{EPolarization} which was adopted for
the low-energy model.\cite{antispiral}
The vector of polarization is
parallel to the ${\bf b}$ axis, in agreement with the experiment.\cite{Taniguchi}
Then, the value of $P_b$ can be estimated as $2.0$, $3.8$, and $4.4$ $\mu$C/m$^2$ for
$q_a$$=$ $-$$1/3$, $-$$1/4$, and $-$$1/5$, respectively. Thus, $P_b$ is small, but it is
well consistent with the fact that the DM interactions are also small and
can be regarded as a small perturbation to the
homogeneous spin-spiral texture, formed by isotropic exchange interactions.
Nevertheless, we would like to emphasize that the
DM interactions are essential for deforming the spin spiral, breaking the inversion
symmetry, and producing the finite value of $P_b$.

  The experimental polarization is an order of magnitude larger: $P_b \sim  50 \mu$C/m$^2$.\cite{Taniguchi}
However, it can be still regarded as a `small value' in comparison with many other
multiferroic systems.\cite{MF_review} Therefore, it is quite consistent with our main idea that
$P_b$ is a result of small perturbation of the magnetic structure, caused by the relativistic effects.

  There may be several reason why the
experimental value of $P_b$ is larger than the theoretical one:

  (i) The experimental $P_b$ may also include
some lattice effects, in response to lowering of the magnetic symmetry;\cite{Shanavas}

  (ii) A unusual aspect of MnWO$_4$ is that the inversion symmetry is broken in the
``high-temperature'' phase, while the low-temperature phase remains centrosymmetric
(for comparison, the situation in perovskite manganites is exactly the opposite).\cite{MF_review}
Thus, in addition to the FE polarization, one should find some mechanism, which would stabilize
the phase AF2 over the phase AF1 at finite $T$. This mechanism may involve some temperature effects.

  (iii) The low-energy model is designed for the semi-quantitative analysis.
It is not always possible to expect a good quantitative agreement with the experimental data, because some
ingredients can be missing in the model. In principle, the GGA$+$$U$ calculations are also of a
semi-quantitative level, because the value of $P_b$ depends on the adjustable parameter $U$.\cite{remark1}
In our model analysis, the small value of $P_b$ may be again related to the overestimation of
isotropic exchange interactions (see Fig.~\ref{fig.J}). Then, the deformation of the spin-spiral
texture, caused by the relativistic effects, is underestimated. Therefore, $P_b$ is also
underestimated. From this point of view, the behavior of isotropic exchange interactions,
the total energies, and the FE polarization is consistent with each other,
and the main efforts towards quantitative description of the FE polarization and the phase
diagram of MnWO$_4$ should be concentrated on the correct description of isotropic exchange interactions.

\section{\label{sec:summary} Summary}

  In summary, we have provided the microscopic explanation for the origin of
FE activity in MnWO$_4$.
The multiferroicity in this compound is caused by the conflict of DM and
isotropic exchange interactions in the AF2 state.
Thus, MnWO$_4$ is multiferroic not simply because of the spiral structure.
It is essential to have conflicting interactions, which deform the spin spiral,
break the inversion symmetry and, thus, give rise to the FE activity.

  \textit{Acknowledgements}.
This work is partly supported by the grant of the Ministry of
Education and Science of Russia N 14.A18.21.0889.

\appendix

\section{\label{sec:appendix}
Nonexistence of ferroelectricity in homogeneous spin-spiral state without
spin-orbit interaction}

\begin{figure}[h!]
\begin{center}
\includegraphics[width=15cm]{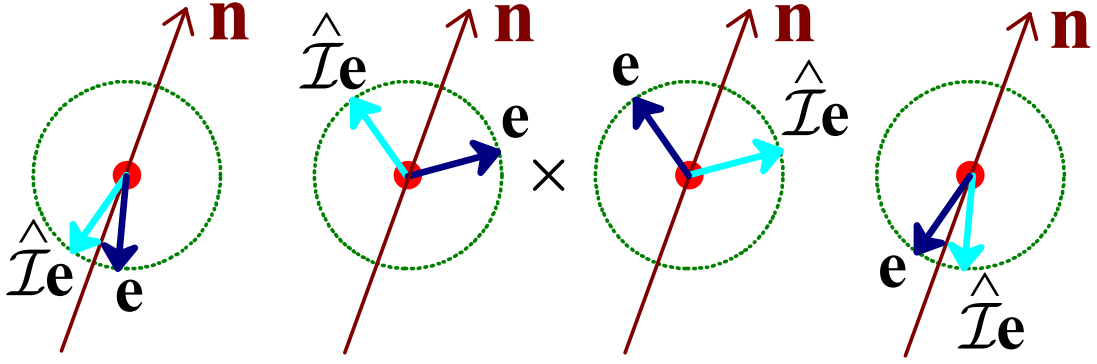}
\end{center}
\caption{\label{fig.cartoon}
(Color online) Cartoon picture, explaining how the inverted spin spiral can be transformed
to the original one by the uniform rotation of spins. The inversion center is marked by cross.
The directions of magnetic moments in the original spin spiral (${\bf e}$) are shown by the
dark (blue) vectors, and those in the inverted spin spiral ($\hat{\cal I}{\bf e}$) are
shown by the light (cyan) vectors. The inverted spin structure can be transformed to the original one
by $180^{\circ}$ rotation of spins around the axis ${\bf n} || ({\bf e}$$+$$\hat{\cal I}{\bf e})$,
which has the same direction at all sites of the lattice.
}
\end{figure}

  In this appendix, we will show that, although the spatial inversion is not the
symmetry operation of the homogeneous spin-spiral state, it can be always combined with an
appropriate rotation spins, which transforms the inverted spin structure to the original one.
The ferroelectric polarization in such a situation will be equal to zero. The prove is extremely simple,
but first we would like to illustrate the basic idea on a cartoon picture for the one-dimensional
spin spiral (see Fig.~\ref{fig.cartoon}).
The basic property of the spin spiral is such that: if ${\bf e}$ is the direction of
spin at certain magnetic site and $\hat{\cal I}{\bf e}$ is the direction of spin at the same site
after the inversion of the lattice,
one can always define the axis ${\bf n} || ({\bf e}$$+$$\hat{\cal I}{\bf e})$, which will have
the same direction at all magnetic sites. Then, the inverted spin structure can be
transformed to the original one by the uniform $180^{\circ}$ rotation of spins around this axis.

  Now, we will provide a more general prove of this statement.
To be specific, consider the situation, realized in MnWO$_4$
and assume that there are two sites in the primitive cell, which are located as $\boldsymbol{\tau}$ and
$-$$\boldsymbol{\tau}$, respectively. Thus, these two cites can be transformed to each other by the
inversion operation $\hat{\cal I}$ and the inversion center is located in the origin.
Then, the translations $\boldsymbol{R}$ will specify the location of all other inversion centers,
while the atomic position will be given by the vectors $\boldsymbol{R}$$\pm$$\boldsymbol{\tau}$.

  In the spin-spiral texture, the directions of magnetic moments are given by
\begin{equation}
{\bf e}_{\boldsymbol{R} \pm \boldsymbol{\tau}} =
\left(
\begin{array}{c}
\cos \bf{q} \cdot ( \boldsymbol{R} \pm \boldsymbol{\tau} ) \\
\sin \bf{q} \cdot ( \boldsymbol{R} \pm \boldsymbol{\tau} ) \\
0 \\
\end{array}
\right).
\label{eqn:e}
\end{equation}
For our purposes, it is sufficient to consider the situation where all spins rotate in the
$\boldsymbol{ab}$ plane. The generalization for an arbitrary orientation of the rotation plane
is straightforward and will be discussed at the end of this appendix.

  Consider the inversion around an arbitrary center $\boldsymbol{R}_0$. It transforms an
arbitrarily taken site $\boldsymbol{R}$$\pm$$\boldsymbol{\tau}$ to
$\boldsymbol{R}_0$$-$$(\boldsymbol{R}$$\pm$$\boldsymbol{\tau}$$-$$\boldsymbol{R}_0) = 2\boldsymbol{R}_0$$-$$\boldsymbol{R}$$\mp$$\boldsymbol{\tau}$.
Then, the direction of spin at the site $2\boldsymbol{R}_0$$-$$\boldsymbol{R}$$\mp$$\boldsymbol{\tau}$ will change from
${\bf e}_{2\boldsymbol{R}_0 - \boldsymbol{R} \mp \boldsymbol{\tau}}$
to ${\bf e}_{\boldsymbol{R} \pm \boldsymbol{\tau}}$.
Thus, our goal is to find a transformation $\hat{\cal R}$, which would rotate ${\bf e}_{\boldsymbol{R} \pm \boldsymbol{\tau}}$ back to
${\bf e}_{2\boldsymbol{R}_0 - \boldsymbol{R} \mp \boldsymbol{\tau}}$. This transformation is the $180^{\circ}$ rotation
around the axis ${\bf n} || ({\bf e}_{2\boldsymbol{R}_0 - \boldsymbol{R} \mp \boldsymbol{\tau}}$$+$${\bf e}_{\boldsymbol{R} \pm \boldsymbol{\tau}})$.
The corresponding transformation matrix is given by
$$
\hat{\cal R} =
\left(
\begin{array}{ccc}
\phantom{-}\cos (2 \bf{q} \cdot \boldsymbol{R}_0 ) & \phantom{-}\sin (2 \bf{q} \cdot \boldsymbol{R}_0 ) & \phantom{-}0 \\
\phantom{-}\sin (2 \bf{q} \cdot \boldsymbol{R}_0 ) & - \cos (2 \bf{q} \cdot \boldsymbol{R}_0 ) & \phantom{-}0 \\
0 & 0 & -1 \\
\end{array}
\right).
$$
Since $\hat{\cal R}$ does not depend on $\boldsymbol{R}$, it corresponds to the uniform rotation of all spins.
Without SO coupling, this is a transformation to the equivalent magnetic state. Thus, although $\hat{\cal I}$
is formally broken, $\hat{\cal R}\hat{\cal I}$ is the symmetry operation, which transforms the
spin spiral to itself. Since the uniform rotation of spins $\boldsymbol{R}$ does not affect the polarization, the existence of the
symmetry operation $\hat{\cal R}\hat{\cal I}$ means that the ferroelectric activity in the
homogeneous spins-spiral state
is forbidden.

  This is a general property of the spin spiral: although we have considered a specific lattice geometry, which is
more relevant to MnWO$_4$, absolutely the same argument can be repeated, for example, for orthorhombic manganites, where the
magnetic sites are located in the centers of inversion.

  For an arbitrary orientation of the spin rotation plane,
${\bf e}_{\boldsymbol{R} \pm \boldsymbol{\tau}}$ in Eq.~(\ref{eqn:e}) should be replaced by
$\hat{\cal D}{\bf e}_{\boldsymbol{R} \pm \boldsymbol{\tau}}$, where $\hat{\cal D}$ is just another uniform rotation of spins,
specifying the orientation of this plane. Then, $\hat{\cal R}$ should be replaced
$\hat{\cal D} \hat{\cal R} \hat{\cal D}^{-1}$. After that all arguments can be repeated again.

\end{document}